\documentclass[aps,pra,reprint,superscriptaddress,floatfix,showpacs]{revtex4-1}
\usepackage{bm}
\usepackage{amsmath,amssymb} 
\usepackage{graphicx}

\usepackage[usenames]{color}

\begin{document}

\title{Josephson plasma oscillations and the Gross-Pitaevskii equation:
\\
Bogoliubov approach \textit{vs} two-mode model }

\author{Alessia Burchianti}
\affiliation{LENS European Laboratory for Non-linear Spectroscopy, and Dipartimento di Fisica e Astronomia - Universit\`a di Firenze, I-50019 Sesto Fiorentino, Italy}
\affiliation{CNR-INO Istituto Nazionale di Ottica, I-50019 Sesto Fiorentino, Italy}

\author{Chiara Fort}
\affiliation{LENS European Laboratory for Non-linear Spectroscopy, and Dipartimento di Fisica e Astronomia - Universit\`a di Firenze, I-50019 Sesto Fiorentino, Italy}
\affiliation{CNR-INO Istituto Nazionale di Ottica, I-50019 Sesto Fiorentino, Italy}

\author{Michele Modugno}
\affiliation{\mbox{Depto. de F\'isica Te\'orica e Hist. de la Ciencia, Universidad del Pais Vasco UPV/EHU, 48080 Bilbao, Spain}}
\affiliation{IKERBASQUE, Basque Foundation for Science, 48011 Bilbao, Spain}

\begin{abstract}
We show that the Josephson plasma frequency for a condensate in a double-well potential, whose dynamics is described by the Gross-Pitaevskii (GP) equation, can be obtained with great precision by means of the usual Bogoliubov approach, whereas the two-mode model - commonly constructed by means of a linear combinations of the low-lying states of the GP equation - generally provides accurate results only for weak interactions. A proper two-mode model in terms of the Bogoliubov functions is also discussed, revealing that in general a two-mode approach is formally justified only for not too large interactions, even in the limit of very small amplitude oscillations.
Here we consider specifically the case of a one-dimensional system, but the results are expected to be valid in arbitrary dimensions.
\end{abstract}

\date{\today}


\maketitle

\section{Introduction}

The Josephson effect \cite{josephson1962} is a clear manifestation of the macroscopic quantum coherence existing between two-weakly coupled superfluids/superconductors. Since its discovery, it has been investigated in a wide variety of physical systems including superconductors \cite{barone1982}, superfluid Helium \cite{pereverzev1997,backhaus1997,davis2002,sukhatme2001} and more recently trapped cold atoms \cite{cataliotti2001,albiez2005} and exciton-polaritons in microcavities \cite{abbarchi2013}.
Besides offering a wealth of accessible experimental parameters, such as interactions and particle statistics, ultracold quantum gases can be easily manipulated by means of magnetic and optical potentials. In such systems, a single Josephson junction can be implemented starting from an atomic Bose Einstein condensate (BEC) confined in a double well potential, as originally proposed in Ref. \cite{javanainen1986}. Over the years, many authors have investigated this paradigmatic model, addressing the non trivial effect of interactions both from the theoretical \cite{dalfovo1996,milburn1997,smerzi1997,zapata1998,raghavan1999,giovanazzi2000,zhang2001,sakellari2002,ananikian2006,giovanazzi2008,ichihara2008,julia-diaz2010,jezek2013} and experimental side \cite{cataliotti2001,albiez2005,levy2007,leblanc2011,trenkwalder2016,fattori2016}, also extending its study to fermionic superfluid atomic samples \cite{spuntarelli2007,ancilotto2009,zou2014,valtolina2015}. In addition, also the effects of thermally induced phase fluctuations \cite{gati2006} and of dissipation have been investigated \cite{labouvie2016}.

Within the formalism of the Gross-Pitaevskii (GP) equation for BECs, Josephson plasma oscillations are tipically described by means of a two-mode model \cite{smerzi1997,pitaevskii2016}, where the two-modes $\phi_{L,R}$ are usually taken either as the interacting ground state of the isolated traps corresponding to the left and right wells with $N/2$ particles \cite{smerzi1997,zapata1998,giovanazzi2000}, or as a linear combination of the lowest and first-excited solutions of the GP equation for the whole system \cite{raghavan1999,sakellari2004,danshita2005,ananikian2006,gati2007,adhikari2009,julia-diaz2010,pitaevskii2016}. 
Though these approaches may be good approximations in some cases, in general they are not rigorous, as the first implies an ideal decoupling of the two wells (that are instead linked), and the second implicitly makes use of the superposition principle, that in general is not valid in the presence of nonlinearity. In particular, since the Josephson plasma frequency $\omega_{J}$ is defined in the limit of small amplitude oscillations, it is not justified to construct the left and right modes $\phi_{L,R}$ by making use of the first-excited solutions of the GP equation with order N particles, as only a small fraction of the total number of particles is expected to populate the excited state in that limit. In other words, each mode function (and
hence the eigenfrequencies) of the GP equation depends on the number of
particles in that mode and so a barely excited mode is a different
object from a fully excited state.
Then, in general one may expect the usual two-mode model to provide accurate predictions for $\omega_{J}$ when the interaction energy does not exceed the kinetic energy (Rabi regime \cite{leggett2001}), but not necessarily in the opposite limit, namely in the proper Josephson regime. Indeed, the fact that the two-mode model can be inaccurate in reproducing the correct value of $\omega_{J}$ has already been pointed out by some authors \cite{ananikian2006,japha2011,jezek2013,valtolina2015}.

In the present paper, motivated by the recent experiment \cite{valtolina2015} that has explored the Josephson plasma oscillations for molecular BECs with large interactions,  we present a systematic analysis of the solutions of the GP equation by means of a two-mode Bogoliubov approach, finding an excellent agreement for any value of the interactions.
This approach is justified by the fact that the Bogoliubov theory correctly describes the GP dynamics in case of small oscillations around the ground-state solution \cite{dalfovo1999}. In fact, the relevance of the Bogoliubov theory in describing the physics of the Josephson effect has been already discussed by a number of authors \cite{paraoanu2001,meier2001,danshita2005,japha2011}. In particular, in \cite{meier2001} the Bogoliubov approach was used to calculate the Josephson current between two weakly interacting BECs that are spatially separated by a tunnel barrier.
In \cite{paraoanu2001} the authors demonstrated that the quanta of the two-mode Josephson Hamiltonian $H=-E_{J}/N (a^{\dagger}b+ab^{\dagger}) + E_{c}/4((a^{\dagger}a)^{2} + (b^{\dagger}b)^{2})$ are in fact the Bogoliubov excitations of the same Hamiltonian.
In Ref. \cite{danshita2005} it was considered a weakly interacting BEC in a box-shaped double-well potential, and it was showed, by varying the barrier height, that the crossover from the dipole mode to the Josephson plasma mode occurs in the lowest energy excitation of the Bogoliubov spectrum. In that paper the Bogoliubov frequency for the first excited mode was found to be in agreement with that calculated with the usual prescription in terms of a linear combination of the lowest and first-excited solutions of the Gross-Pitaevskii equation \cite{raghavan1999,sakellari2004,danshita2005,ananikian2006,gati2007,adhikari2009,pitaevskii2016}. In \cite{japha2011}, a different approach - based on an approximation of the Bogoliubov theory - was proposed: the $\phi_{L,R}$ are constructed from a linear combination of the ground-state solution $\psi_{0}$, and the first excited state $\psi_{1}$ of the GP equation with the mean field term generated by $\psi_{0}$. The effects of higher modes have also been considered \cite{julia-diaz2010,gillet2014}.

Our work present a systematic comparison of the results directly obtained by solving the GP equation with the ones obtained by using the Bogoliubov approach or a standard two-mode model.
This analysis reveals that the usual approach for constructing the two-mode basis functions \cite{raghavan1999,sakellari2004,danshita2005,ananikian2006,gati2007,adhikari2009,pitaevskii2016}, rapidly becomes inaccurate \cite{jezek2013} as the interactions are increased. Moreover, we show that a proper two-mode model for describing the Josephson plasma oscillations, constructed by means of the Bogoliubov approach, reveals that in general the two-mode approach is formally justified only for weak interactions, and that by increasing interactions it eventually breaks down, even in the limit of very small amplitude oscillations. Here we consider specifically the case of a one-dimensional system, but the results are expected to be valid in arbitrary dimensions.

The outline of the paper is the following: in Sec. \ref{sec:model} we introduce the GP model we shall use throughout this work, along with the usual two-mode model considered in the literature (Sec. \ref{sec:2mode}) and the standard Bogoliubov formalism (Sec. \ref{sec:bogol}) for describing the dynamics of excitations in the linear regime. Consistently with the two-mode picture, only the ground state and the first excited Bogoliubov modes are considered here. Then, in Sec. \ref{sec:results} we present a systematic comparison between the numerical solution of the GP equation and the two approaches just mentioned above. In particular, we show that the oscillation frequency obtained from the Bogoliubov approach perfectly matches the GP result, whereas the two-mode model generally fails in reproducing the correct results. Also, in Sec. \ref{sec:2mode} we discuss whether the Bogoliubov theory justifies the formulation of the problem in terms of a two left and right modes, finding that it cannot be formally justified for arbitrary interactions. Finally, conclusions are drawn in Sec. \ref{sec:concl}.

\section{Model}
\label{sec:model}

Let us consider a (quasi) one-dimensional condensate of $N$ particles with mass $m$ confined in a double well potential $V_{dw}(x)$,
\begin{equation}
V_{dw}(x) \equiv \frac12m\omega_{x}^{2}x^{2} + V_{0}{e}^{-2x^{2}/w^{2}},
\end{equation}
($w<\sqrt{4V_{0}/(m\omega^{2})}$) whose dynamics is described by the following Gross-Pitaevskii equation
\begin{equation}
i\hbar\partial_{t}\psi(x,t)= \left[\hat{H}_{0} +u_{0}|\psi(x,t)|^{2} \right]\psi(x,t)
\end{equation}
with
\begin{equation}
\hat{H}_{0}=-\frac{\hbar^2}{2M}\nabla_{x}^{2}+V_{dw}(x)
\end{equation}
and $u_{0}=gN$, $g$ being the one-dimensional interaction strength, and the condensate wave function being normalized to unity, $\int \! dx \ |\psi(x)|^2 = 1$. 

The above equation can be conveniently written in dimensionless form, for example by expressing all the quantities in oscillator units (e.g. energies in units of $\hbar\omega_{x}$, lengths in units of $a_{x}=\sqrt{\hbar/m\omega_{x}}$)
\begin{equation}
i\partial_{\tilde{t}}\tilde{\psi}=\left[-\frac12\nabla_{\tilde{x}}^{2} + \frac12\tilde{x}^{2} + \tilde{V}_{0}{e}^{-2\tilde{x}^{2}/\tilde{w}^{2}}
+\tilde{u}_{0}|\tilde{\psi}|^2\right]\tilde{\psi} 
\label{eq:gpe-dimensionless}
\end{equation}
where $\tilde{\psi}=\sqrt{a_{x}}\psi$, $\tilde{u}_{0}=u_{0}/(a_{x}\hbar\omega_{x})$, $\tilde{t}=\omega_{x}t$, and $\nabla_{\tilde{x}}^{2}=\partial^{2}_{\tilde{x}}$. For simplicity, in the following the tilde will be omitted.

\subsection{Two-mode model}
\label{sec:2mode}

The two-mode approximation - that here we review for the sake of clarity, and for fixing the notations - consists in assuming that the condensate wave function can be written as 
\begin{equation}
\psi(x,t)=c_{L}(t)\psi_{L}(x)+c_{R}(t)\psi_{R}(x)
\label{eq:twomode}
\end{equation}
where the functions $\psi_{L,R}$ are localized in the left and right well, have unit norm, and are orthogonal ($\langle\psi_{L}|\psi_{R}\rangle=0$). Then the GP equation 
(\ref{eq:gpe-dimensionless}) can be transformed into a set of equations for the two coefficients $c_{\alpha}(t)$ ($\alpha=L,R$). By inserting (\ref{eq:twomode}) in (\ref{eq:gpe-dimensionless}), left multiplying by $\psi_{L}$, integrating over space, and retaining all possible terms \cite{ananikian2006}, one gets
\begin{align}
\label{eq:cdot2}
i\hbar\dot{c}_{L}(t)&=
c_{L}(t) (E_{0L} +|c_{L}(t)|^{2}U_{LLLL}) - c_{R}(t){\cal K}_{LR}
\\
&+2c_{L}(t)Re[c_{R}^{*}(t)c_{L}(t)]U_{LLRL}
\nonumber\\ 
&+c_{L}(t)|c_{R}(t)|^{2}U_{LRRL}
+
c_{R}(t)|c_{L}(t)|^{2}U_{L LLR}
\nonumber\\
&+2c_{R}(t)Re[c_{R}^{*}(t)c_{L}(t)]U_{L LRR}
\nonumber\\
&+c_{R}(t)|c_{R}(t)|^{2}U_{L RRR} 
\nonumber
\end{align}
where we have defined
\begin{align}
E_{0\alpha}&\equiv\langle\psi_{\alpha}|\hat{H}_{0}|\psi_{\alpha}\rangle=\int dx\psi_{\alpha}(x)\hat{H}_{0} \psi_{\alpha}(x)
\\
{\cal K}_{\alpha\beta}&\equiv-\langle\psi_{\alpha}|\hat{H}_{0}|\psi_{\beta}\rangle=-\int dx\psi_{\alpha}(x)\hat{H}_{0}\psi_{\beta}(x)
\\
U_{\alpha mn\beta}&\equiv u_{0}\int dx \psi_{\alpha}(x)\psi_{m}(x)\psi_{n}(x)\psi_{\beta}(x)
\end{align}
Similarly, the equation for $c_{R}$ is obtained by exchanging $L$ with $R$ in the former expression. 
Then, by defining
\begin{equation}
c_{\alpha}(t)=\sqrt{N_{\alpha}(t)}e^{i\phi_{\alpha}(t)}
\label{eq:twomode-c}
\end{equation}
and 
\begin{align}
\phi&\equiv\phi_{L}-\phi_{R}
\label{eq:phasediff}
\\
z&\equiv N_{L}-N_{R}
\label{eq:zeta}
\end{align}
with $N_{L}+N_{R}=1$ \footnote{The number of atoms in each well is given by $N\times N_{\alpha}$.}, one eventually gets
\begin{equation}
\begin{cases}
\hbar\dot{z} &= 2\left({\cal K} -U_{\alpha\alpha\alpha\beta}\right)\sqrt{1-z^{2}}\sin\phi
\\
&-\displaystyle(1-z^{2}) U_{\alpha\alpha\beta\beta}\sin2\phi
\\
\hbar \dot{\phi}&= -(U_{\alpha\alpha\alpha\alpha}- 2U_{\alpha\alpha\beta\beta})z 
\\
&-2\left({\cal K} -U_{\alpha\alpha\alpha\beta}\right)\frac{z}{\sqrt{1-z^{2}}}\cos\phi+zU_{\alpha\alpha\beta\beta}\cos2\phi
\end{cases}
\nonumber
\end{equation}

In the limit of small oscillations, $z\ll1$, $\phi\ll1$, the above equations reduce to
\begin{equation}
\begin{cases}
\hbar\dot{z} &\simeq 2\left({\cal K} -U_{\alpha\alpha\alpha\beta}- U_{\alpha\alpha\beta\beta}\right)\phi
\\
\hbar \dot{\phi}&\simeq -\left(U_{\alpha\alpha\alpha\alpha} -3U_{\alpha\alpha\beta\beta} + 2{\cal K} -2U_{\alpha\alpha\alpha\beta}\right)z
\end{cases}
\end{equation}
corresponding to harmonic oscillations of the population imbalance $z$ with  frequency
\begin{widetext}
\begin{equation}
\omega_{J}^{fTM}=\frac{1}{\hbar}\sqrt{2\left({\cal K} -U_{\alpha\alpha\alpha\beta}- U_{\alpha\alpha\beta\beta}\right)
\left(\displaystyle U_{\alpha\alpha\alpha\alpha} + 2{\cal K} -2U_{\alpha\alpha\alpha\beta}-3U_{\alpha\alpha\beta\beta} \right)}.
\label{eq:omega-ftm}
\end{equation}
\end{widetext}
In the following, this approach will be referred to as the \textit{full} two-mode (fTM) model.
Instead, the usual two-mode approximation which neglects the terms $U_{\alpha\alpha\alpha\beta}$ and $U_{\alpha\alpha\beta\beta}$ consists in taking
 \begin{equation}
\omega_{J}^{TM}=\frac{1}{\hbar}\sqrt{2{\cal K}(2{\cal K}+U_{\alpha\alpha\alpha\alpha})}
\label{eq:omega-tm}
\end{equation}
which, for small values of $2{\cal K}$ gives  $\omega_{J}^{TM}=\sqrt{2{\cal K}U_{\alpha\alpha\alpha\alpha}}/\hbar$, corresponds to the usual formula $\omega_{J}=\sqrt{E_{c}E_{J}}/\hbar$ \cite{pitaevskii2016}, with $E_{J}={\cal K}N$, and $E_{c}=2U_{\alpha\alpha\alpha\alpha}/N$. Instead, for vanishing interactions ($U_{\alpha\alpha\alpha\alpha}=0$), Eq. (\ref{eq:omega-tm}) gives $\omega_{J}^{noint}={2{\cal K}}/{\hbar} $,
corresponding to the energy difference between the first excited state and the ground state of the linear Schr\"odinger equation.

With these notations, the Rabi regime is characterized by $2U_{\alpha\alpha\alpha\alpha}/{\cal K}\ll1$, whereas the Josephson regime holds in the opposite case, $2U_{\alpha\alpha\alpha\alpha}/{\cal K}\gg1$ \footnote{
In order to facilitate the reader, here we list the correspondence between the notations of Ref. \cite{leggett2001} and our dimensionless notations: $K\to 2U_{\alpha\alpha\alpha\alpha}/N$, $E_{J}\to N{\cal K}$.}.
We notice that in general one can define an additional Fock regime dominated by quantum fluctuations, for $2U_{\alpha\alpha\alpha\alpha}/{\cal K}\gg N^{2}$ \cite{leggett2001}, that is obviously beyond the scope of any meanfield theory, as it is the present case. 

\subsection{Bogoliubov approach}
\label{sec:bogol}

Instead of the ansatz (\ref{eq:twomode}), in the limit of small oscillations (linear regime) one can simply use the standard Bogoliubov approach, namely ($\mu$ is the condensate chemical potential)
\begin{equation}
\psi(x,t)=e^{-i\mu t/\hbar}\left[\psi_{0}(x)+\delta\psi(x,t)\right]
\label{eq:expansion1}
\end{equation}
that, by defining
\begin{equation}
\hat{\cal L}\equiv \hat{H}_{0} +g\psi_{0}^{2}-\mu\,,
\end{equation}
corresponds to the following set of equations for the fluctuations $\delta\psi$
\begin{equation}
i\hbar\partial_{t}
\left(\begin{matrix}
\delta\psi\\
\delta\psi^{*}\\
\end{matrix}
\right)=\left(\begin{matrix}
 \hat{\cal L}&g\psi_{0}^{2}\\
-g\psi_{0}^{2}&-\hat{\cal L}\\
\end{matrix}
\right)\left(\begin{matrix}
\delta\psi\\
\delta\psi^{*}\\
\end{matrix}
\right)\equiv \hat{\cal L}_{B}
\left(\begin{matrix}
\delta\psi\\
\delta\psi^{*}\\
\end{matrix}
\right).
\end{equation}
Then, by expanding the column vector $(\delta\psi,\delta\psi^{*})$ in terms of eigenmodes of $\hat{\cal L}_{B}$ (Bogoliubov modes) as
\begin{equation}
\left(\begin{matrix}
\delta\psi(x,t) \\
\delta\psi^{*}(x,t) \\
\end{matrix}
\right)=\sum_{k}
c_{k}(t)\left(\begin{matrix}
u_{k}(x)\\
v_{k}(x)\\
\end{matrix}
\right)+
c^{*}_{k}(t)\left(\begin{matrix}
v^{*}_{k}(x)\\
u^{*}_{k}(x)\\
\end{matrix}
\right)
\end{equation}
with
\begin{equation}
\hat{\cal L}_{B}\left(\begin{matrix}
u_{k}(x)\\
v_{k}(x)\\
\end{matrix}
\right)=\varepsilon_{k}
\left(\begin{matrix}
u_{k}(x)\\
v_{k}(x)\\
\end{matrix}
\right),
\end{equation}
one gets
\begin{equation}
i\hbar\dot{c}_{k}(t)=\varepsilon_{k}c_{k}(t)\,.
\end{equation}

Consistently with the hypothesis (\ref{eq:twomode}), let's now assume that only the lowest Bogoliubov mode, with energy $\varepsilon=\hbar\omega_{B}$, is actually populated
\begin{align}
\delta\psi(x,t)&=c(t)u(x) + c^{*}(t)v^{*}(x)
\label{eq:expansion2}
\end{align}
with
\begin{equation}
c(t)=c_{0}e^{-i\omega_{B}t}\,,
\end{equation}
and 
\begin{equation}
c_{0}=\int dx \Psi_{0}(x)\left[u^{*}(x)-v^{*}(x)\right]
\label{eq:c0}
\end{equation}
in the representation where the $u$ and $v$ functions are orthogonal to $\psi_{0}$ \cite{morgan1998,blakie2002,castin2001}, and $\Psi_{0}(x)$ being the initial wave function (that we assume to be real). We recall that $u$, and $v$ obey to the standard normalization condition $\int\!\! dx \left(|u(x)|^{2} - |v(x)|^{2}\right)=1$ \cite{dalfovo1999}.
We also assume $|c(t)|^{2}=|c_{0}|^{2}\ll1$, corresponding to the fact that in the limit of small oscillations, only a small fraction of the total particles occupy the excited state. Then, we can write
\begin{equation}
\psi(x,t) = \psi_{0}(x) + c_{0}\left[u(x)e^{-i\omega_{B}t} + v(x)e^{+i\omega_{B}t}\right],
\label{eq:bogol-exp}
\end{equation}
where, for symmetry reasons, both functions $u(x)$ and $v(x)$ are antisymmetric (see later on), and can be chosen real, without loss of generality. 
The total density of particles is therefore 
\begin{align}
 n(x,t)&= |\psi_{0}(x) + \delta\psi(x,t)|^{2}
 \label{eq:dens}
\\
&\simeq
|\psi_{0}(x)|^{2}
+2c_{0}\cos(\omega_{B}t)\psi_{0}(x)\left[u(x) + v(x)\right]
\nonumber
\end{align}
where we have discarded terms of order $c_{0}^{2}$.
The expression for $n(x)$ is a linear combination of a symmetric and an antisymmetric term (respectively the first and the second term),
describing an oscillation of the particle occupation of the left and right well, with frequency $\omega_{B}$. 
In fact, by integrating the former expression over the positive or negative $x$ semi-axis, and taking into account the symmetries of the problem we can write
\begin{equation}
N_{L,R}(t)=A\pm B\cos(\omega_{B}t)
\end{equation}
where 
\begin{align}
A&=\int_{0}^{+\infty}\!\!\!\!\!\!\!\!dx|\psi_{0}(x)|^{2},
\quad
B=2c_{0}\int_{0}^{+\infty}\!\!\!\!\!\!\!\!dx\psi_{0}(x)\left[u(x) + v(x)\right]
\end{align}
so the population imbalance, defined as $z=(N_{L}-N_{R})/(N_{L}+N_{R})$ (see Eq. (\ref{eq:zeta})), oscillates as
\begin{equation}
z(t)=2B\cos(\omega_{B}t).
\label{eq:omega-b}
\end{equation}
We remark that this results follows rather straightforwardly from the Bogoliubov expansion and the symmetries of the system. In the following section we shall see that the Bogoliubov approach indeed describes very accurately the Josephson plasma oscillations in the linear regime (small amplitude oscillations).

\section{Numerical results}
\label{sec:results}

In this section we compare the predictions of  Eq. (\ref{eq:omega-ftm}), (\ref{eq:omega-tm}) with the numerical solution of the Gross-Pitaevskii equation in Eq. (\ref{eq:gpe-dimensionless}), and the Bogoliubov frequency $\omega_{B}$, by varying the interaction parameter $u_0$. For illustration purposes, here we choose ${w}=0.3$ and ${V}_{0}=50$ (unless otherwise stated), that correspond to a double well configuration, within reach of current experiments (see e.g. \cite{valtolina2015}).
Nevertheless, we remark that the general results of the following analysis are independent of these specific values.
As for the interaction strength $u_{0}$, that is the only free parameter left (see Eq. (\ref{eq:gpe-dimensionless})), we vary it in the range $[1,400]$ in order to cover a wide spectrum of the Josephson regime (see later on). This choice corresponds to an intermediate interacting regime, where the applicability of the GP theory is well justified \footnote{In general, the interactions can be very large and yet the system can be still weakly interacting (e.g. in the Thomas-Fermi regime), so that the GP theory still applies; see also the discussion on page 474 of Ref. \cite{dalfovo1999}.}. For example, for a typical case of an elongated condensate of $10^4$ $^{87}$Rb atoms with trapping frequency $f_r=100$ Hz (radial), $f_z=10$ Hz (axial) (where the use of an effective one-dimensional approach can be appropriate and the mean-field GP equation has been tested in a wide range of experiments), the value of the reduced 1d coupling constant is $u_0=g_{3D}/(3\pi a_{r}^{2})=3\cdot10^2$, $a_r=\sqrt{\hbar/(m 2\pi f_r)}$ being the radial oscillator length, ($u_0=3\cdot10^3$ for $f_r=1$ kHz). The same considerations apply also in the presence of a barrier, as Josephson oscillations are essentially long-wavelength ``classical'' excitations that can be well described by the GP theory, even in the Thomas-Fermi regime (see e.g. \cite{dalfovo1996,zapata1998,valtolina2015}).

We prepare the initial state as the ground state $\psi_{0}(x)$ of the double well potential, and we compute the corresponding Bogoliubov spectrum, for different values of the interaction constant $u_{0}$ \footnote{The stationary GP equation is solved by imaginary time evolution \cite{dalfovo1999}. The time-dependent GP equation is solved by means of a split-step method that makes use of fast Fourier transforms \cite{jackson1998}. The Bogoliubov equations are transformed into matrix equations by projection in Fourier space \cite{modugno2004}, and then diagonalized with LAPACK routines (available at www.netlib.org/lapack).}.
\begin{figure}
\centerline{\includegraphics[width=0.9\columnwidth]{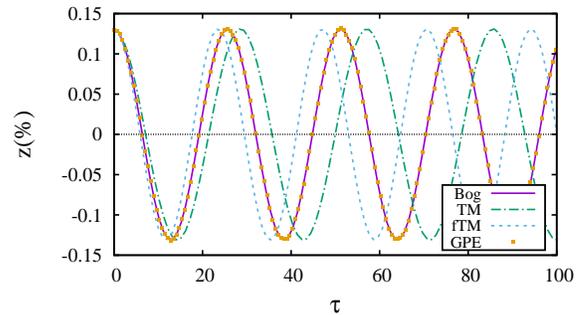}}
\caption{(Color online) Oscillations of the population imbalance $z(t)$, for $u_{0}=16$ ($\mu/V_{0}=0.1$).
}
\label{fig:gpe}
\end{figure}
Then, at time $t=0$ we trigger the dynamics by suddenly displacing the potential by a (small) fixed distance $\delta x=0.002$, in order to guarantee that the system remains in the linear regime (harmonic oscillations) in the whole range of interactions considered here. This corresponds to an initial population imbalance $z_{0}$ of the order of $0.1\%$. 
In Fig. \ref{fig:gpe} we show the first oscillations of the population imbalance $z(t)$ as obtained from the GP equation (points), along with the harmonic oscillation with frequencies $\omega_{J}^{fTM}$, $\omega_{J}^{TM}$, and $\omega_{B}$ (for the same initial imbalance), corresponding to the predictions of the various approaches discussed in the previous section (see Eqs. (\ref{eq:omega-ftm}), (\ref{eq:omega-tm}), and (\ref{eq:omega-b}), respectively), for $u_{0}=16$ ($\mu/V_{0}=0.1$). This figure shows that the prediction of the Bogoliubov approach perfectly fits with the GP solution, whereas the usual two-mode model deviates significantly, regardless of the approximation used.

\begin{figure}
\centerline{\includegraphics[width=0.9\columnwidth]{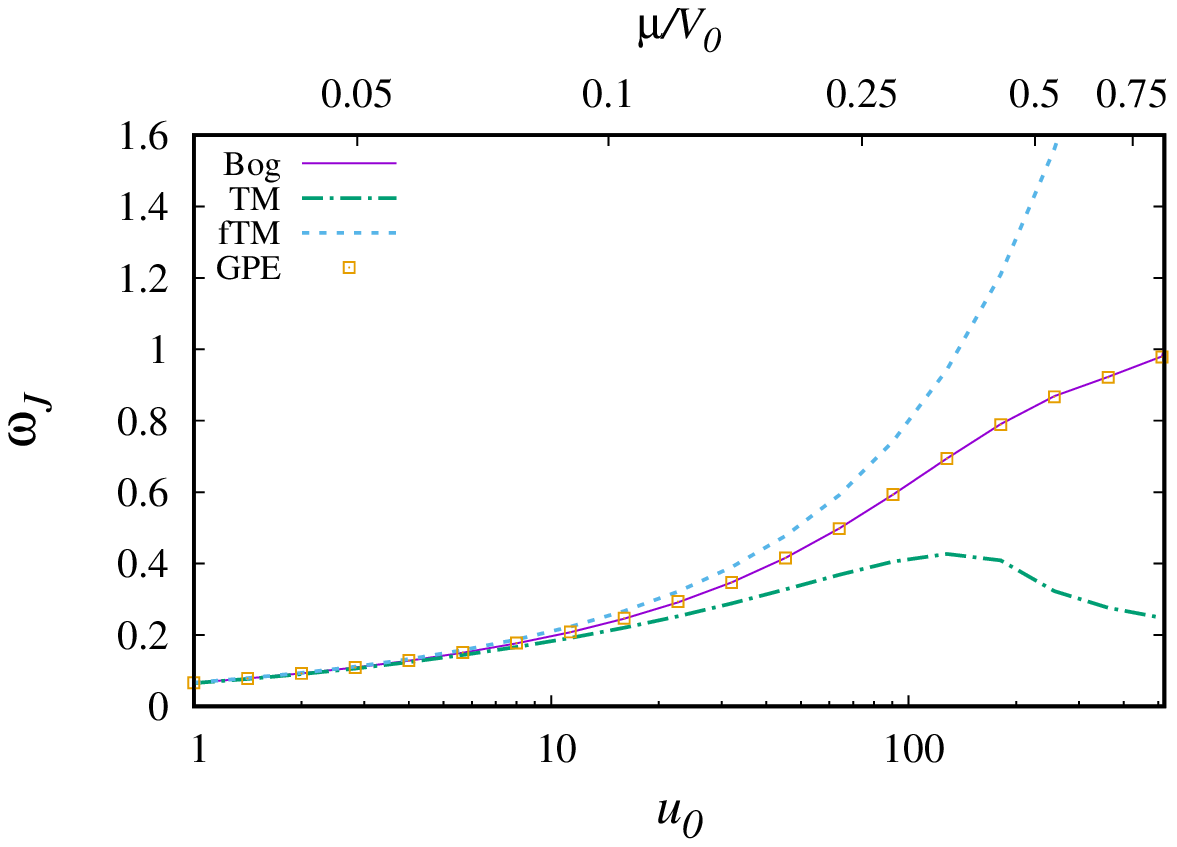}}
\caption{(Color online) Josephson frequency as obtained from the different approaches discussed in the text (Eq. (\ref{eq:omega-ftm}), (\ref{eq:omega-tm}), and the Bogoliubov frequency $\omega_{B}$), compared with the results of the GP equation (points), as a function of $u_{0}$, for $V_{0}=50$. The upper $x$-axis shows the corresponding value of of $\mu/V_{0}$ (that is a monotonic increasing function of $u_{0}$).
}
\label{fig:omegaj}
\vspace{0.5cm}
\centerline{\includegraphics[width=0.9\columnwidth]{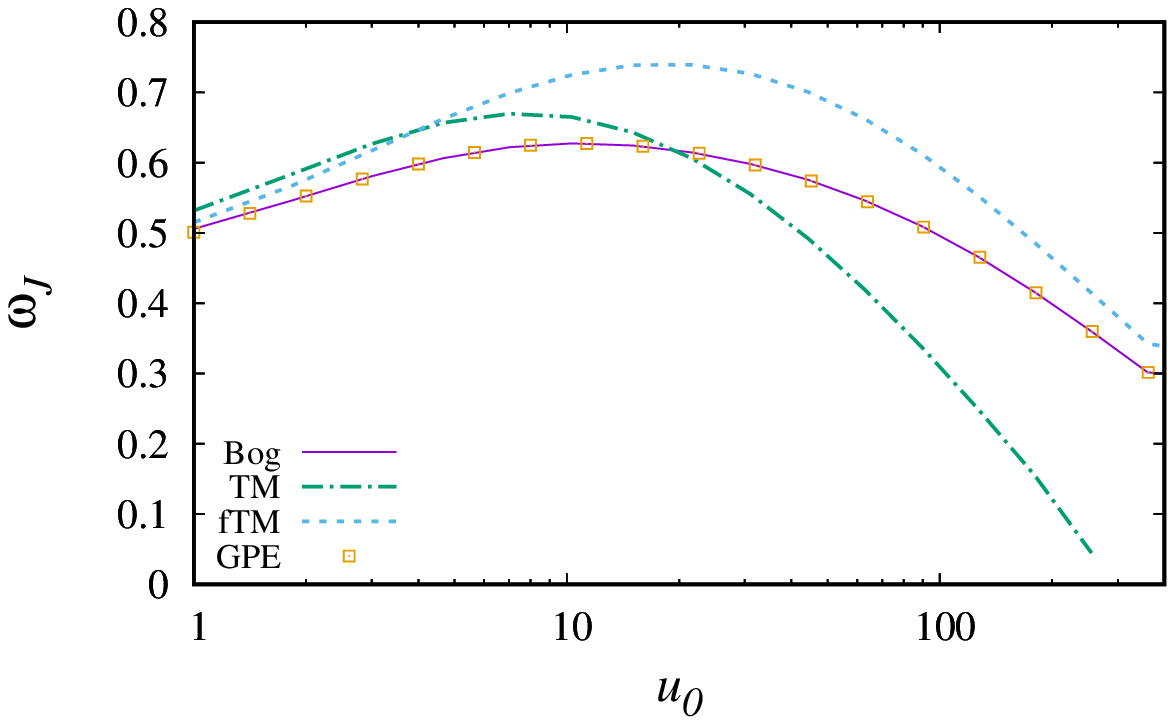}}
\caption{(Color online) Josephson frequency as obtained from the different approaches discussed in the text, compared with the results of the GP equation (points), as a function of $u_{0}$ for fixed $\mu/V_{0}=0.25$.
}
\label{fig:omegaj2}
\end{figure}

In order to provide a comprehensive overview of the behavior of the system as a function of the interaction, in Fig. \ref{fig:omegaj} and \ref{fig:omegaj2} we plot the Josephson frequency as obtained from the different approaches discussed in the text, compared with the results of the GP equation (points), for increasing interactions. The latter is obtained by fitting the oscillations of $z(t)$ with a function $f_{z}(t)\equiv A\cos(\omega t)$, with $A$ and $\omega$ as fitting parameters.
 In Fig. \ref{fig:omegaj} we keep the barrier intensity fixed to $V_{0}=50$, and we vary the interaction parameter $u_{0}$ (that, in this case, corresponds to $3\cdot10^{2}\lesssim 2U_{\alpha\alpha\alpha\alpha}/{\cal K}\lesssim 3\cdot10^{5}$, deeply in the Josephson regime). For convenience, we also show the values of the ratio $\mu/V_{0}$ (upper $x$-axis), that is a monotonic increasing function of $u_{0}$. In this case, the failure of the two-mode approach for relatively large values of $u_{0}$ can be attributed both to the fact that the increase of the interactions makes the use of the first-excited solution of the GP equation more and more inaccurate, and also because the system eventually exits the weak coupling regime, as the tunneling increases with $\mu/V_{0}$.
 Then, in order to focus on the former mechanism, in Fig. \ref{fig:omegaj2} we show the same quantities, again as a function of $u_{0}$, but at a fixed ratio $\mu/V_{0}=0.25$  ($V_{0}$ here is varied along with $u_{0}$, in order to keep the ratio $\mu/V_{0}$ fixed). 
In this case the first point in the graph lies close to the boundary between Rabi and Josephson regimes, but then the system rapidly enters the Josephson regime ($5\lesssim2U_{\alpha\alpha\alpha\alpha}/{\cal K}\lesssim5\cdot10^{6}$).

These figures show that the Bogoliubov frequency perfectly matches the frequency extracted from the GP equation in all the range of interactions considered here, whereas the usual two-mode model is 
reliable only for small values of $u_{0}$, reflecting the fact that a proper treatment of small amplitude oscillations in an interacting system requires the use of the Bogoliubov approach, as expected.

\section{Two-mode model \textit{a la} Bogoliubov}
\begin{figure*}[t!]
\centerline{\includegraphics[width=0.8\columnwidth]{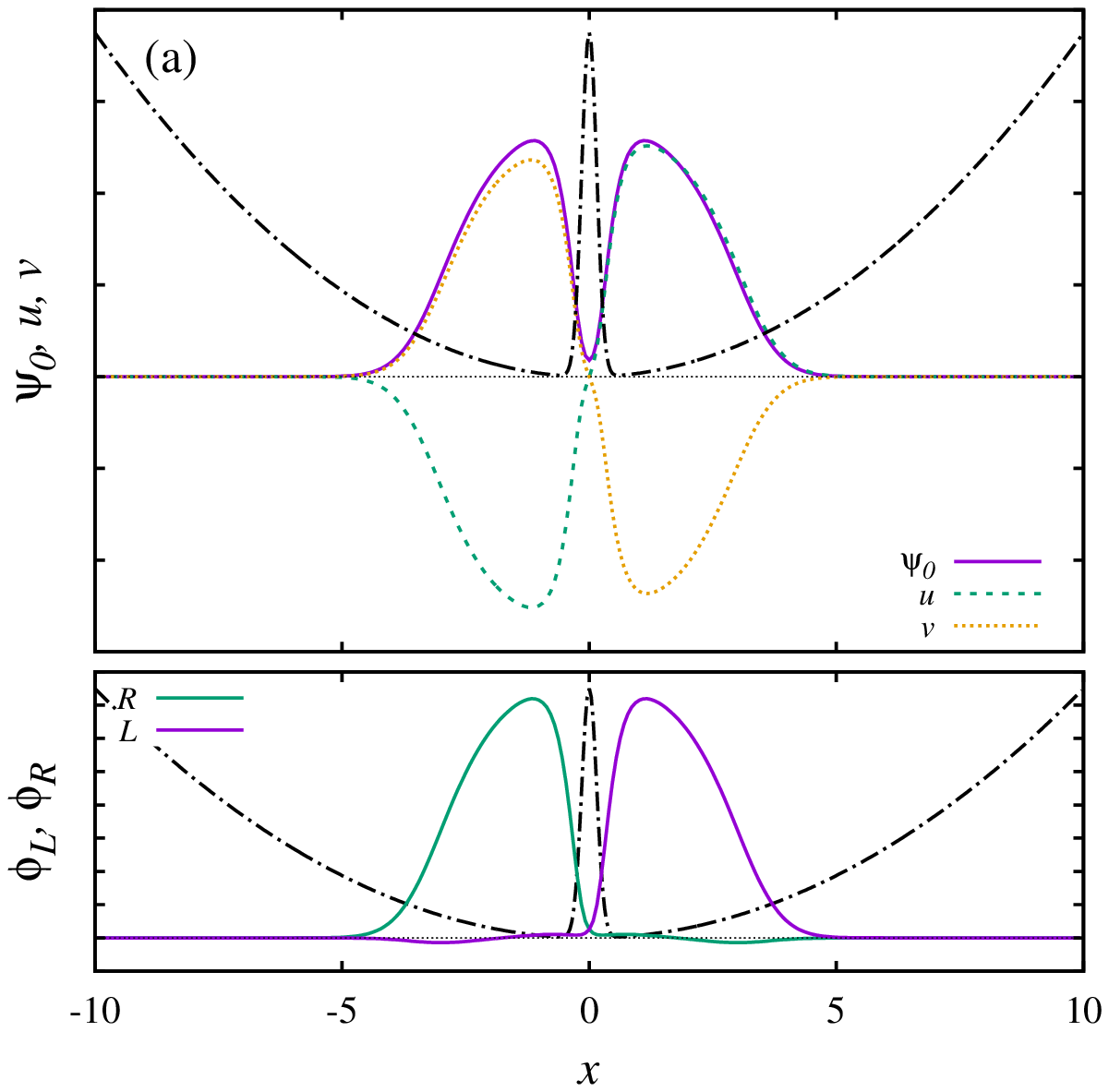}\hspace{2cm}
\includegraphics[width=0.8\columnwidth]{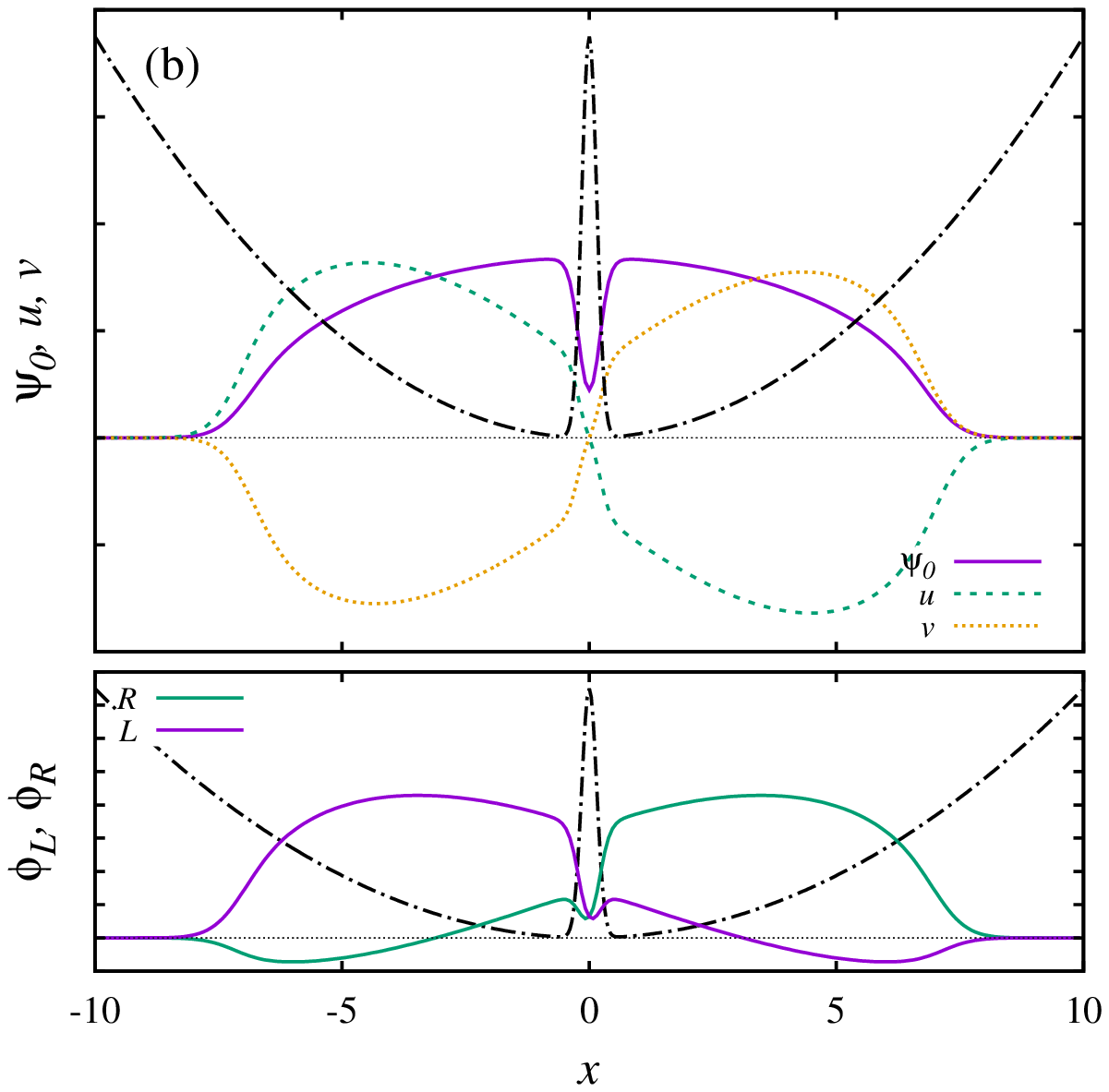}}
\caption{(Color online) (top) Plot of the ground state wave function $\psi_{0}(x)$, and of the Bogoliubov functions $\tilde{u}(x)$ and $\tilde{v}(x)$, for (a) $\mu/V_{0}=0.1$ ($u_{0}=16$) and (b) $\mu/V_{0}=0.5$ ($u_{0}=220$). (bottom) Left and right states obtained from symmetric and antisymmetric combination of $\psi_{0}(x)$ and $\tilde{u}(x)$, see Eq. (\ref{eq:sacombination}). In both panels, the double well potential is also shown.}
\label{fig:grounds}
\end{figure*}

In this section we shall discuss whether the Bogoliubov frequency $\omega_{B}$ shown in Figs. \ref{fig:omegaj} and \ref{fig:omegaj2} can be related to a two-mode model with the left and right basis functions defined by means of the functions entering the Bogoliubov expansion. We anticipate that this is so only for not too large interactions, where the shape of the $u$ and $v$ functions still resemble that of the ground state $\phi_{0}$. Conversely, as $u_{0}$ is increased, the interactions distinctively affect the shape of the $u$ and $v$ functions in the tail region (see later on), so that the \textit{three} functions entering the Bogoliubov expansion in Eqs. (\ref{eq:expansion1}), (\ref{eq:expansion2}) cannot be rewritten in terms of just two basis functions.

Let us start by recalling that in the noninteracting limit the $v$ components of the solutions with positive energy (positive norm \cite{castin2001}) are vanishing, so that the standard decomposition in terms of the ground state $\psi_{0}$ and the first excited state $u$,
\begin{equation}
\phi_{L,R}(x) = \frac{1}{\sqrt{2}}(\psi_{0}(x)\pm u(x)),
\label{eq:sacombination}
\end{equation}
yields a good basis of functions localized in the left and right well. Essentially, this decomposition is possible when the portion of each function to be decomposed has the same shape of the basis functions $\phi_{L,R}$, modulo a scale factor. Notice also that in general one has $\phi_{L}(x)=\phi_{R}(-x)$, owing to the symmetries of the problem.

When interactions are present, the expression in Eq. (\ref{eq:bogol-exp}) contains three functions - namely $\psi_{0}$, $u$, and $v$ - and in general it is not obvious that such an expression can be projected onto a basis of left and right functions, as the shape of the three functions may be affected differently by the interactions. In particular we shall see that for large interactions, though $v(x)\simeq -u(x)$, their shape can be quite different from that of $\psi_{0}$.

However, we notice that for not too large interaction the expression in Eq. (\ref{eq:sacombination}) still yields a good basis of functions localized in the left and right well, provided that we change the normalization of the Bogoliubov functions ($u,v\rightarrow \tilde{u},\tilde{v})$ as $\int dx |\tilde{u}(x)|^{2}=1$, $\int dx |\tilde{v}(x)|^{2}=1-1/N_{u}= N_{\tilde{v}}$ (where we have introduced the following notation: $N_{f}\equiv\int dx|f(x)|^{2}$). This is shown in Fig. \ref{fig:grounds}a, where we plot the ground state wave function $\psi_{0}(x)$, the Bogoliubov functions $\tilde{u}(x)$ and $\tilde{v}(x)$, and the left and right basis functions $\phi_{L,R}$, along with the double well potential, for $u_{0}=16$.
In this regime, Eq. (\ref{eq:sacombination}) should be used along with the following expression for $\tilde{v}$,
\begin{equation}
\tilde{v}(x)=\frac{\alpha}{\sqrt{2}}\left( \psi_{L}(x) - \psi_{R}(x)\right),
\label{eq:vprojected}
\end{equation}
with $\alpha/\sqrt{2}=\langle\tilde{v}|\psi_{L}\rangle=-\langle\tilde{v}|\psi_{R}\rangle$, $\alpha^{2}=N_{\tilde{v}}$. Then, from Eqs. (\ref{eq:bogol-exp}), (\ref{eq:sacombination}), and (\ref{eq:vprojected}) one gets
\begin{align}
\psi(x,t) &= \psi_{0}(x) + \tilde{c}_{0}\left[\tilde{u}(x)e^{-i\omega_{B}t} + \tilde{v}(x)e^{+i\omega_{B}t}\right]
\nonumber
\\
&= \frac{1}{\sqrt{2}}\psi_{L}(x)\left[1 + \tilde{c}_{0}(e^{-i\omega_{B}t} + \alpha e^{+i\omega_{B}t})\right]\nonumber\\
& 
+ \frac{1}{\sqrt{2}}\psi_{R}(x)\left[1 - \tilde{c}_{0}(e^{-i\omega_{B}t} + \alpha e^{+i\omega_{B}t})\right],
\label{eq:twomodeb}
\end{align}
%
with $\tilde{c}_{0}=N_{u}\int dx\Psi_{0}(x)[\tilde{u}^{*}(x)-\tilde{v}^{*}(x)]$ (see Eq. (\ref{eq:c0})).
Then, comparing the previous expression with Eq. (\ref{eq:twomode-c}), one has
\begin{align}
c_{L,R}(t)= \frac{1}{\sqrt{2}}&\left[1 \pm \tilde{c}_{0}(1 + \alpha)\cos(\omega_{B}t) \right.
\nonumber\\
&\quad\left.\mp i\tilde{c}_{0}(1 - \alpha)\sin(\omega_{B}t)\right]
\end{align}
so that the total number on each side of the barrier, $N_{L,R}(t)=|c_{L,R}(t)|^{2}$, is 
\begin{align}
N_{L,R}(t)\simeq\frac{1}{2}&\left[1 
+\tilde{c}_{0}(1 - \alpha)\sin(\omega_{B}t)
\right.\nonumber\\
&\left.\qquad\pm 2\tilde{c}_{0}(1+\alpha)\cos(\omega_{B}t)\right],
\end{align}
where quadratic terms in $\tilde{c}_{0}$ have been discarded, consistently with the assumption in Eq. (\ref{eq:bogol-exp}).
The previous expression implies that the population imbalance oscillates with frequency $\omega_{B}$, namely
\begin{equation}
z(t)=2\tilde{c}_{0}(1+\alpha)\cos(\omega_{B}t),
\end{equation}
that is exactly the Bogoliubov frequency already shown and compared with the other methods in Figs. \ref{fig:omegaj}
and \ref{fig:omegaj2} (see also Eq. (\ref{eq:omega-b})). Similarly, one can compute the phase difference as (see Eq. (\ref{eq:phasediff}))
 \begin{equation}
\phi(t)=\phi_{L}-\phi_{R}\simeq-2\tilde{c}_{0}(1 - \alpha)\sin(\omega_{B}t),
\end{equation}
corresponding again to sinusoidal oscillations at the plasma frequency, with a phase shift of $\pi/2$ with respect to $z(t)$, as expected for a Josephson plasma oscillation.

This approximate picture breaks down for higher values of the interactions, where one may still have $v(x)\simeq-u(x)$, but the combination in Eq. (\ref{eq:sacombination}) no longer provides functions localized in the left and right well. As anticipated, this is due to the fact that the shape of $u$ and $v$ is quite different from that of $\psi_{0}$. An example is shown in Fig. \ref{fig:grounds}b, obtained for $u_{0}=220$, where it is evident that the interactions  strongly modify the shape of the $u$ and $v$ functions in the tail region, with respect to that of the ground state $\psi_{0}$. Then, if one tries to construct the functions $\phi_{L/R}$ according to Eq. (\ref{eq:sacombination}), they would not be localized in one of two wells (see bottom panel of Fig. \ref{fig:grounds}b). As a matter of fact, this implies that the \textit{three} functions entering the Bogoliubov expansion in Eqs. (\ref{eq:expansion1}), (\ref{eq:expansion2}) cannot be rewritten in terms of just two basis functions.

These results show that not only the standard two-mode model (in any of its versions) provides inaccurate results in many regimes, but also that in general a two-mode approach is formally justified only for not too large interactions.

\section{Conclusions}
\label{sec:concl}

We have shown that the frequency of the Josephson plasma oscillations for a condensate in a double well (within the Gross-Pitaevskii theory) corresponds to the Bogoliubov frequency of the lowest excited mode, for arbitrary values of the interactions. This contrasts to the prediction of the usual two-mode approach - in terms of linear combinations of the low-lying states of the Gross-Pitaevskii equation - that is reliable only in the weak link regime, for low values of the interactions. These results have been found by means of a systematic analysis of the Gross-Pitaevskii equation and the Bogoliubov equations as a function of the interactions. They confirm some previous analyses performed at fixed values of the interaction, by different authors \cite{paraoanu2001,meier2001,danshita2005,japha2011,jezek2013}.
 In addition, we have shown that the Bogoliubov approach provides a proper formalism for defining a two-mode model, 
also revealing that in general the two-mode approach is justified only for weak interactions, and that it eventually breaks down by increasing interactions
, even in the limit of very small amplitude oscillations. Though we have considered specifically the case of a one-dimensional system, the general results obtained here are expected to be valid in arbitrary dimensions. Moreover, we expect the analysis in terms of the Bogoliubov modes to be extremely effective also in the nonlinear regime \cite{morgan1998,blakie2002}, when the system exits from the plasma oscillations regime. This will be the object of a future work.

\textit{Acknowledgments.}
We acknowledge useful discussions with M. Fattori, G. Roati, and G. Barontini.
This work has been supported by the UPV/EHU under program UFI 11/55, the Spanish Ministry of Science and Innovation and the European Regional Development Fund FEDER through Grant No. FIS2015-67161-P (MINECO/FEDER), and the Basque Government through Grant No. IT-472-10. A. B. was supported by European Research Council grant no. 307032 QuFerm2D.


%

\end{document}